\documentclass{article}
\usepackage{authblk}
\usepackage{graphicx}
\usepackage{dcolumn}
\usepackage{cite}
\usepackage{amsmath}
\usepackage[numbers,sort&compress]{natbib}
\usepackage{amssymb}
\usepackage{rotating}
\usepackage{caption}
\usepackage{color}
\title{Exploring the unknown territory of Dromions of (2+1) dimensional Generalized Nonlinear Schr\"odinger Equation} 
\author[1]{C. Senthil Kumar}
\author[2*]{R.Radha}
\affil[1]{Department of Physics, Vinayaka Mission’s Kirupananda Variyar Engineering College, Vinayaka Mission’s Research Foundation (DU), Salem  636 308, India}
\affil[2]{Centre for Nonlinear Science (CeNSc), Post-Graduate and Research Department of Physics, Government College for Women (Autonomous), Kumbakonam 612 001, India,}
\affil[*]{Correspondong Author: vittal.cnls@gmail.com}
\begin{document}
\maketitle

\begin{abstract}
In this paper, we travel  through an unknown territory of dromions,  to unearth and show the new properties/attributes of "dromions" which have never been brought to the fore since they were first discovered by Boiti etal\cite{Boiti88}. These new attributes are brought out
by investigating a generalized (2+1) dimensional Nonlinear Schr\"odinger (NLS) equation by exploiting Truncated Painlev\'e approach. The signatures that can be attributed to dromions include ``existence of firewall and reflection at the boundary, uneven distribution of energy among different bound states, amplitude dependence on adjacent dromions, etc.". We have corroborated the main analytical results with numerical simulations also.  We categorically state that these properties are universal and can be extracted in any (2+1) dimensional nonlinear partial differential equation. We do believe that these properties which shed more light on the behaviour of dromions may have wider repercussions in nonlinear optics, Bose-Einstein condensates and plasma physics.
\end{abstract}

The identification of localized excitations in (2+1) dimensions  such as \cite{Santini90, Estevez99, Lou00, Mihalache21,rizvi2025abundant,aims2025solitary,mehanna2025comprehensive}  rogue waves \cite{Kharif2003603,  Zakharov2006677, Dysthe2008287, Osborne2009}, lumps \cite{Ablowitz19782180,Estevez20077213,Albares20172305,Wu20191755, cw1, cw3}, dromions \cite{Boiti88, Fokas90,Radha94,Lou19957227}, etc, which  find applications in different branches of science such as nonlinear optics \cite{Boyd08}, Bose-Einstein 
\noindent Condensates \cite{Bludov09}, fluid dynamics \cite{Apel07}, plasma physics \cite{Kono10}, electron-acoustic waves \cite{Ghosh13} and  magnetized plasma \cite{saini16} has completely revolutionized our understanding of integrability in higher dimensions. It is quite well known that the higher dimensional nonlinear  dynamical systems exhibit richness in their solutions. This richness can be attributed to the existence of lower dimensional arbitrary functions of space and time.  The recent application of the concepts of Machine Learning(ML) \cite{LI2024108242,FU2023115771} and Physics Informed Neural Networks(PINN) \cite{epjp2025bilinear,ZHOU2023164} to higher dimensional nonlinear partial differential equations has enabled researchers to dig deep into the dynamics of higher dimensional integrable systems. 
The investigations in other frontier areas of science like epidemology\cite{raza2025mathematical}, biological transport, ecological systems and population dynamics \cite{ahmed2025iterative} has kept the investigation of (2+1) dimensional nonlinear dynamical systems alive even today.

It should be pointed out that among the above mentioned nonlinear localized excitations,  lumps,  are algebraically localized and do not interact with each other
while rogue waves arise from nowhere and disappear without a trace and are found to be unstable.

On the other hand, it was   Boiti et al\cite{Boiti88} who came up with exponentially localized solutions in two spatial dimensions which they termed as "dromions" derived from the Greek word "dromos" which means "boundaries". They are exponentially localized solutions decaying in both 
$x$ and $y$ directions.  They essentially arise by virtue of the intersection/interaction of two non-parallel ghost solitons. Boiti et al pointed out that they can be  driven any where in the two dimensional plane by choosing the lower dimensional boundaries or arbitrary functions of space and time  properly. They went on to add that dromions can interact both elastically and inelastically \cite{Boiti88, Fokas90, Hietarinta90, Radha94, Lai99, Tang02}.
It should be pointed out that the dynamics of dromions cannot yet be regarded as fully understood. Although the nuances behind the collisional  dynamics of dromions  in (2+1) dimensional nonlinear pdes are well documented, they still manifest several intricate and compelling features that remain to be fully unearthed.  The fact that there are abundant lower dimensional arbitrary functions of space and time in (2+1) dimensional nonlinear partial differential equations encourages researchers to look for a deeper investigation of dromions. 

In this paper, to explore the unknown territory of dromions, we focus our attention on  a sample (2+1) dimensional nonlinear pde, namely, the  generalized nonlinear Schr\"odinger (NLS) equation introduced by Fokas \cite{Fokas94}, which integrates  Davey-Stewartson-I (DS I) and Davey-Stewartson-III (DS III) equations. Recently, Cheng et al have obtained solitons, rogue wave solutions for DSI equation \cite{CHENG20251306}. For nonlocal DSIII equation, Wang et al have constructed solitons, breathers and lumps using Kadomtsev–Petviashvili (KP) hierarchy reduction technique \cite{WANG2024107739}. The (2+1) dimensional Davey-Stewartson-I equation \cite{Davey74} finds application in hydrodynamics, nonlinear optics \cite{Zakharov84}, plasma physics \cite{Nishinari93}, etc. This system is shown to admit Painlev\'e Property \cite{Ganesan87, Leo88} and admits localized solutions such as dromions \cite{Boiti88,Fokas90,Tang02}.
To solve this system, we employ Truncated Painlev\'e Expansion method \cite{Senthil05} to obtain closed form solutions involving  lower dimensional arbitrary functions and generate higher order solutions.

The  paper is organized as follows:  In section 2, we exploit the Truncated Painlev\'e Expansion method for solving the (2+1) dimensional generalized NLS equation. We then obtain  its closed form solutions in terms of lower dimensional  arbitrary functions of space and time.
Then, by exploiting  these arbitrary functions of space and time suitably, we construct higher order dromions in section 3 and study their collisional dynamics to bring out their interesting attributes in section 4. Finally, we conclude with a summary of the results. 

\section{(2+1) dimensional generalized Nonlinear Schr\"o- dinger Equation and Truncated Painlev\'e Approach}
We employ Truncated Painlev\'e Expansion method for the  (2+1) dimensional generalized NLS equation  \cite{Fokas94} of the following form

\begin{align}
iq_t-(\alpha-\beta)q_{\xi\xi}+(\alpha+\beta)q_{\eta\eta}  -2 \lambda q \bigg[(\alpha+\beta) \bigg( \int_{-\infty}^{\xi}|q|_{\eta}^2 d \xi'+u_1(\eta,t) \bigg) &  \nonumber 
\\ 
-(\alpha-\beta)\bigg(\int_{-\infty}^{\eta}|q|_{\xi}^2 d\eta'+u_2(\xi,t) \bigg)\bigg]&=0.\label{GNLS}
\end{align}

In Eq.~(\ref{GNLS}), $\lambda = \pm 1$, $\alpha, \beta$ are real constants, while $u_1(\eta,t), u_2(\xi,t)$ are arbitrary real valued functions. For the parametric choices $\alpha=0, \beta=1$ and $\alpha=1, \beta=0$,  Eq.~(\ref{GNLS}) reduces to DSI and DSIII respectively. For the choice $\alpha=\beta$, $u_1=u_2=0$, Eq.~(\ref{GNLS}) reduces to the simplest scalar equation.
By introducing potentials $U$ and $V$, eq.~(\ref{GNLS}) can be rewritten  as
\begin{subequations}
\begin{eqnarray}
iq_t-(\alpha-\beta)q_{\xi\xi}+(\alpha+\beta)q_{\eta\eta}
-2 \lambda q [(\alpha+\beta)V-(\alpha-\beta)U]&=&0,\\
V_{\xi} &=& |q|^2_\eta, \\
U_{\eta} &=& |q|^2_\xi.
\end{eqnarray}  \label{c2_P1}
\end{subequations}
Rewriting Eq.~(\ref{c2_P1}) by substituting $q=a$ and $q^*=b$, Eq. (\ref{c2_P1}) becomes
\begin{subequations}
\begin{eqnarray}
ia_t-(\alpha-\beta)a_{\xi\xi}+(\alpha+\beta)a_{\eta\eta} -2 \lambda a [(\alpha+\beta)V-(\alpha-\beta)U]=0, \\
-ib_t-(\alpha-\beta)b_{\xi\xi}+(\alpha+\beta)b_{\eta\eta} -2 \lambda b [(\alpha+\beta)V-(\alpha-\beta)U]=0, \\
V_{\xi} = (ab)_\eta, \\
U_{\eta} = (ab)_\xi.
\end{eqnarray}  \label{c2_P2}
\end{subequations}
We express the variables $a, b, V$ and $U$ in terms of the Laurent series, 
\begin{subequations}
\begin{eqnarray}
a=\sum_{j=0}^{\infty} a_j \phi^{j-1}, \\
b=\sum_{j=0}^{\infty} b_j \phi^{j-1}, \\
V=\sum_{j=0}^{\infty} V_j \phi^{j-2}, \\
U=\sum_{j=0}^{\infty} U_j \phi^{j-2},
\end{eqnarray}  \label{c2_P3}
\end{subequations}
where $\phi$ is the noncharacteristic singular manifold. Eq.(\ref{c2_P1}) can be shown to admit the P-property and the manifold $\phi$ can be proven to be arbitrary. 
We then  truncate the above Laurent series at the constant level term to obtain :
\begin{subequations}
\begin{eqnarray}
a=\frac{a_0}{\phi}+a_1, \\
b=\frac{b_0}{\phi}+b_1, \\
V=\frac{V_0}{\phi^2}+\frac{V_1}{\phi}+V_2, \\
U=\frac{U_0}{\phi^2}+\frac{U_1}{\phi}+U_2.
\end{eqnarray}  \label{c2_P4}
\end{subequations}
Now, assuming the seed solution satisfying  eq.(\ref{c2_P2}), we have
\begin{subequations}
\begin{eqnarray}
a_1=b_1=0, \;\;\; \\
V_2=V_2(\eta,t),\;\;\; U_2=U_2(\xi,t),
\end{eqnarray}  \label{c2_P5}
\end{subequations}
To invoke  the truncated Painlev\'e approach, we now substitute the transformations given by eq.(\ref{c2_P4}) along with the seed solution into Eq.~(\ref{c2_P2}) and collect the coefficients of different powers of $\phi$. 
We start from the lowest powers of $\phi$. 
Collecting  the coefficients of  ($\phi^{-3}, \phi^{-3}, \phi^{-3}, \phi^{-3}$), we have  
\begin{subequations}
\begin{eqnarray}
-2 (\alpha-\beta) a_0 \phi_\xi^2+2 (\alpha+\beta) a_0 \phi_\eta^2 
-2  \lambda a_0 [(\alpha+\beta)V_0-(\alpha-\beta)U_0] = 0,\\
-2 (\alpha-\beta) b_0 \phi_\xi^2+2 (\alpha+\beta) b_0 \phi_\eta^2 
-2 \lambda b_0 [(\alpha+\beta)V_0-(\alpha-\beta)U_0] = 0, \\
-2V_0 \phi_\xi = -2a_0 b_0 \phi_\eta, \\
-2U_0 \phi_\eta = -2 a_0 b_0 \phi_\xi.
\end{eqnarray}  \label{c2_P6}
\end{subequations}
Solving the above equations, we obtain the leading order coefficients as
\begin{equation}
\lambda V_0 = \phi_\eta^2, \;\;\; \lambda U_0 = \phi_\xi^2, \;\;\; \lambda a_0 b_0 = \phi_\xi \phi_\eta.  \label{c2_P7}
\end{equation}
Next, collecting the coefficients of  ($\phi^{-2}, \phi^{-2}, \phi^{-2}, \phi^{-2}$), we obtain
\begin{subequations}
\begin{eqnarray}
-i a_0 \phi_t-(\alpha-\beta)(-2 a_{0\xi}\phi_\xi-a_0 \phi_{\xi\xi}) 
+(\alpha+\beta)(-2a_{0\eta}\phi_\eta-a_0 \phi_{\eta\eta}) \nonumber \\
-2 \lambda a_0[(\alpha+\beta)V_1-(\alpha-\beta)U_1]= 0, \label{c2_P8a}\\
i b_0 \phi_t-(\alpha-\beta)(-2 b_{0\xi}\phi_\xi-b_0 \phi_{\xi\xi}) 
+(\alpha+\beta)(-2b_{0\eta}\phi_\eta-b_0 \phi_{\eta\eta}) \nonumber \\
-2 \lambda b_0[(\alpha+\beta)V_1-(\alpha-\beta)U_1]=0, \label{c2_P8b}\\
V_{0\xi}-V_1 \phi_\xi = (a_0b_0)_\eta, \label{c2_P8c}\\
U_{0\eta}-U_1 \phi_\eta = (a_0b_0)_\xi. \label{c2_P8d}
\end{eqnarray}  \label{c2_P8}
\end{subequations}
From Eq.~(\ref{c2_P8c}), we get
\begin{equation}
V_1 = \frac{\phi_{\xi\eta}\phi_\eta-\phi_\xi\phi_{\eta\eta}}{\lambda \phi_\xi}. \label{c2_P9}
\end{equation}
Similarly from Eq.~(\ref{c2_P8d}), we get
\begin{equation}
U_1 = \frac{\phi_{\xi}\phi_{\xi\eta}-\phi_{\xi\xi}\phi_{\eta}}{\lambda \phi_\eta}. \label{c2_P10}
\end{equation}
Substituting Eqs.~(\ref{c2_P7}) in Eq.~(\ref{c2_P8a}), we get
\begin{eqnarray}
    -i\phi_t-2\bigg(\frac{\phi_{\xi\xi}\phi_\eta+\phi_\xi\phi_{\xi\eta}}{\phi_\xi\phi_\eta}-\frac{b_{0\xi}}{b_0}\bigg)\phi_\xi-\phi_{\xi\xi} && \nonumber \\-2\bigg(\frac{\phi_{\xi\eta}\phi_\eta+\phi_\xi\phi_{\eta\eta}}{\phi_\xi\phi_\eta}-\frac{b_{0\eta}}{b_0}\bigg)\phi_\eta 
    -\phi_{\eta\eta}-2\lambda[U_1+V_1]&=&0. \label{t1}
\end{eqnarray}
Substituting for $b_{0\xi}$ and $b_{0\eta}$ from  eq.(\ref{c2_P8b}) in eq.(\ref{t1}), we get
\begin{eqnarray}
-2\bigg(\frac{\phi_{\xi\xi}\phi_\eta+\phi_\xi\phi_{\xi\eta}}{\phi_\xi\phi_\eta}\bigg)\phi_\xi-2\phi_{\xi\xi}
-2\bigg(\frac{\phi_{\xi\eta}\phi_\eta+\phi_\xi\phi_{\eta\eta}} {\phi_\xi\phi_\eta}\bigg)\phi_\eta  \nonumber \\
-2\phi_{\eta\eta}-4\lambda[U_1+V_1]&=&0. \label{t2}
\end{eqnarray}
Now, substituting eq.(\ref{c2_P9}) and eq. (\ref{c2_P10}) in  eq.(\ref{t2}) and simplifying the resultant equation, we obtain a trilinear equation
\begin{equation}
(\phi_\xi^2+\phi_\eta^2) \phi_{\xi\eta} = 0,  \label{c2_P11}
\end{equation}
which implies that $\phi_{\xi\eta}=0$.   From this relation, we can fix the form of the manifold $\phi$ to be 
\begin{equation}
\phi = \phi_1 (\xi,t)+ \phi_2 (\eta,t),  \label{c2_P12}
\end{equation}  
where  $\phi_1(\xi,t)$ and $\phi_2(\eta,t)$ are lower dimensional arbitrary functions of space and time in the indicated variables.
Then, to  solve Eq.~(\ref{c2_P8a}), we consider the following transformation for $a_0$ such that
\begin{equation}
a_0(\xi,\eta,t) = a_{01}(\xi,t) a_{02} (\eta,t).  \label{c2_P13}
\end{equation}
By substituting the above transformation in Eq. (\ref{c2_P8a}), it can be rewritten as
\begin{align}
 a_{02} (-ia_{01} \phi_{1t}-(\alpha-\beta)(-2 a_{01 \xi} \phi_{1 \xi}+a_{01} \phi_{1\xi\xi})) + &  \nonumber \\
a_{01}(-ia_{02} \phi_{2t}+(\alpha+\beta)(-2 a_{02 \eta} \phi_{2 \eta}+a_{02} \phi_{2\eta\eta})) & = 0.  \label{c2_P14}
\end{align}
Splitting the above equation  in the following form 
\begin{subequations}
\begin{eqnarray}
-ia_{01} \phi_{1t}-(\alpha-\beta)(-2 a_{01 \xi} \phi_{1 \xi} 
+a_{01} \phi_{1\xi\xi}) = i \kappa a_{01}, \label{c2_P15a}\\
-ia_{02} \phi_{2t}+(\alpha+\beta)(-2 a_{02 \eta} \phi_{2 \eta} 
+a_{02} \phi_{2\eta\eta}) = -i \kappa a_{02} \label{c2_P15b}
\end{eqnarray}
\end{subequations}
and solving the resultant equations, we obtain
\begin{subequations}
\begin{eqnarray}
a_{01}= F_1(t) \mbox{e}^{\int \frac{-i \phi_{1t}-(\alpha-\beta)\phi_{1\xi\xi}- i \kappa}{-2 \phi_{1\xi}(\alpha-\beta)} d\xi}, \label{c2_P16a} \\
a_{02}= F_2(t) \mbox{e}^{\int \frac{-i \phi_{2t} +(\alpha+\beta)\phi_{2\eta\eta}+ i \kappa}{2 \phi_{2\eta}(\alpha+\beta)} d\eta}, \label{c2_P16b}
\end{eqnarray}\label{c2_P16}
\end{subequations}
where $F_1(t)$ and $F_2(t)$ are arbitrary functions.
Eq. (\ref{c2_P8b}) is also consistent with eq.(\ref{c2_P16}).
Then, collecting the coefficients of ($\phi^{-1}, \phi^{-1}, \phi^{-1}, \phi^{-1}$), we have
\begin{subequations}
\begin{eqnarray}
i a_{0t} -(\alpha-\beta) a_{0\xi\xi}+(\alpha+\beta)a_{0\eta\eta} 
-2 \lambda a_0 [(\alpha+\beta)V_2-(\alpha-\beta)U_2] =0,  \label{c2_P17a}\\
-i b_{0t}-(\alpha-\beta) b_{0\xi\xi}+(\alpha+\beta)b_{0\eta\eta}
-2 \lambda b_0 [(\alpha+\beta)V_2-(\alpha-\beta)U_2] =0, \label{c2_P17b}\\
V_{1\xi}=0, \label{c2_P17c}\\
U_{1\eta}=0. \label{c2_P17d}
\end{eqnarray}  \label{c2_P17}
\end{subequations}
From Eq.~(\ref{c2_P17a}) and  using the earlier results, we obtain
\begin{subequations}
\begin{eqnarray}
U_2&=& -\frac{1}{2 \lambda F_1}\bigg[\frac{i}{(\alpha-\beta)}\bigg(F_{1t}+\int F_1 \bigg[\frac{-i \phi_{1tt}-(\alpha-\beta)\phi_{1\xi\xi t}}{-2 \phi_{1\xi}(\alpha-\beta)}  - \nonumber \\
& & \frac{-i \phi_{1t}-(\alpha-\beta)\phi_{1\xi\xi}-i \kappa}{-2 \phi_{1\xi}^2(\alpha-\beta)} \phi_{1\xi t}\bigg] d\xi \bigg)\nonumber \\
& & -F_1 \bigg( \frac{-i \phi_{1 \xi t}-(\alpha-\beta)\phi_{1\xi\xi\xi}}{-2 \phi_{1 \xi}(\alpha-\beta)}
-\frac{-i \phi_{1t}-(\alpha-\beta)\phi_{1\xi\xi}-i \kappa}{-2\phi_{1\xi}^2(\alpha-\beta)}\phi_{1\xi\xi}+ \nonumber \\
& & \bigg[\frac{-i\phi_{1t}-(\alpha-\beta)\phi_{1\xi\xi}-i \kappa}{-2 \phi_{1\xi}(\alpha-\beta)}\bigg]^2\bigg)\bigg],  \\
V_2&=& \frac{1}{2 \lambda F_2}\bigg[\frac{i}{(\alpha+\beta)}\bigg(F_{2t}+\int F_2 \bigg[\frac{-i \phi_{2tt}+(\alpha+\beta)\phi_{2\eta\eta t}}{2 \phi_{2\eta}(\alpha+\beta)}  -  \nonumber \\
& & \frac{-i \phi_{2t} +(\alpha+\beta)\phi_{2\eta\eta}+i \kappa}{2 \phi_{2\eta}^2(\alpha+\beta)} \phi_{2\eta t}\bigg]d\eta \bigg)\nonumber \\
& & + F_2 \bigg( \frac{-i \phi_{2 \eta t}+(\alpha+\beta)\phi_{2\eta\eta\eta}}{2 \phi_{2 \eta}(\alpha+\beta)}  -\frac{-i \phi_{2t}+(\alpha+\beta)\phi_{2\eta\eta}+i \kappa}{2\phi_{2\eta}^2(\alpha+\beta)}\phi_{2\eta\eta} \nonumber \\
& & +\bigg[\frac{-i\phi_{2t}+(\alpha+\beta)\phi_{2\eta\eta}+i\kappa}{2 \phi_{2\eta}(\alpha+\beta)}\bigg]^2\bigg)\bigg]. 
\label{c2_P18a}
\end{eqnarray}  \label{c2_P18}
\end{subequations}
Eqs. (\ref{c2_P17b}), (\ref{c2_P17c}) and (\ref{c2_P17d}) are consistent with the earlier results.  

Thus, we can write down the solution of the generalized (2+1) dimensional NLS equation as
\begin{subequations}
\begin{eqnarray}
a&=& \frac{F_1 F_2 \sqrt{\phi_{1\xi} \phi_{2\eta}} \mbox{e}^{\int \frac{-i \phi_{1t}-i \kappa}{-2 \phi_{1\xi}(\alpha-\beta)} d\xi}   \mbox{e}^{\int \frac{-i \phi_{2t} + i \kappa}{2 \phi_{2\eta}(\alpha+\beta)}d\eta}}{(\phi_1 (\xi,t)+\phi_2 (\eta,t))},\label{c2_P19a}\\
b&=& \frac{\sqrt{\phi_{1\xi} \phi_{2\eta}}}{\lambda F_1 F_2 \mbox{e}^{\int \frac{-i \phi_{1t}-i \kappa}{2 \phi_{1\xi}(\alpha-\beta)}d\xi}   \mbox{e}^{\int \frac{-i \phi_{2t}+ i \kappa}{2 \phi_{2\eta}(\alpha+\beta)}d\eta}(\phi_1 (\xi,t)+ \phi_2 (\eta,t))}, \label{c2_P19b}\\
V&=&\frac{ \phi_{2\eta}^2}{\lambda (\phi_1 (\xi,t)+ \phi_2 (\eta,t))^2}-\frac{\phi_{2\eta\eta}}{\lambda (\phi_1 (\xi,t)+\phi_2 (\eta,t))}+  \frac{1}{2 \lambda F_2}\bigg[\frac{i}{(\alpha+\beta)}\bigg(F_{2t} \nonumber \\ & &
+\int F_2 \bigg[\frac{-i \phi_{2tt}+(\alpha+\beta)\phi_{2\eta\eta t}}{2 \phi_{2\eta}(\alpha+\beta)}  -  \frac{-i \phi_{2t} +(\alpha+\beta)\phi_{2\eta\eta}+i \kappa}{2 \phi_{2\eta}^2(\alpha+\beta)} \phi_{2\eta t}\bigg]d\eta \bigg) \nonumber \\
& & + F_2 \bigg( \frac{-i \phi_{2 \eta t}+(\alpha+\beta)\phi_{2\eta\eta\eta}}{2 \phi_{2 \eta}(\alpha+\beta)}  -\frac{-i \phi_{2t}+(\alpha+\beta)\phi_{2\eta\eta}+i \kappa}{2\phi_{2\eta}^2(\alpha+\beta)}\phi_{2\eta\eta} \nonumber \\
& & +\bigg[\frac{-i\phi_{2t}+(\alpha+\beta)\phi_{2\eta\eta}+i\kappa}{2 \phi_{2\eta}(\alpha+\beta)}\bigg]^2\bigg)\bigg]. 
\end{eqnarray} 
\begin{eqnarray}
U&=&\frac{\phi_{1\xi}^2}{\lambda ( \phi_1 (\xi,t)+ \phi_2 (\eta,t))^2}-\frac{ \phi_{1\xi\xi}}{\lambda ( \phi_1 (\xi,t)+ \phi_2 (\eta,t))} -\frac{1}{2 \lambda F_1}\bigg[\frac{i}{(\alpha-\beta)}\bigg(F_{1t}
\nonumber\\ 
& &+\int F_1 \bigg[\frac{-i \phi_{1tt}-(\alpha-\beta)\phi_{1\xi\xi t}}{-2 \phi_{1\xi}(\alpha-\beta)} -  \frac{-i \phi_{1t}-(\alpha-\beta)\phi_{1\xi\xi}-i \kappa}{-2 \phi_{1\xi}^2(\alpha-\beta)} \phi_{1\xi t}\bigg] d\xi \bigg) \nonumber \\
 & & -F_1 \bigg( \frac{-i \phi_{1 \xi t}-(\alpha-\beta)\phi_{1\xi\xi\xi}}{-2 \phi_{1 \xi}(\alpha-\beta)}
 -\frac{-i \phi_{1t}-(\alpha-\beta)\phi_{1\xi\xi}-i \kappa}{-2\phi_{1\xi}^2(\alpha-\beta)}\phi_{1\xi\xi}+ 
 \nonumber \\
& &\bigg[\frac{-i\phi_{1t}-(\alpha-\beta)\phi_{1\xi\xi}-i \kappa}{-2 \phi_{1\xi}(\alpha-\beta)}\bigg]^2\bigg)\bigg],
\label{c2_P19d}
\end{eqnarray}  \label{c2_P19}
\end{subequations}
  
Using the condition $a=b^*$ in Eqs.~(\ref{c2_P19a}) and (\ref{c2_P19b}), we obtain the constraint
\begin{equation}
F_1 = \pm \frac{1}{\sqrt{\lambda} F_2}. \label{ }
\end{equation}
Applying the above constraint, the solution given by equation (\ref{c2_P19}) takes the form  (\ref{FS1})-(\ref{FS3}).

Using Painlev\'e Truncation Approach, we can write down the solutions of Eq.~(\ref{c2_P1}) in the following form
\begin{equation}
q = \frac{\pm \sqrt{\phi_{1\xi} \phi_{2\eta}} \mbox{e}^{\int \frac{-i \phi_{1t}-i \kappa}{-2 \phi_{1\xi}(\alpha-\beta)}d\xi}   \mbox{e}^{\int \frac{-i \phi_{2t}+i \kappa}{2 \phi_{2\eta} (\alpha+\beta)}d\eta}}{\sqrt{\lambda}(\phi_1 (\xi,t)+ \phi_2 (\eta,t))} \label{FS1}
\end{equation}
and its squared magnitude takes the form
\begin{equation}
|q|^2= \frac{\phi_{1\xi} \phi_{2\eta}}{\lambda(\phi_1 (\xi,t)+ \phi_2 (\eta,t))^2}. \label{GNLS22} 
\end{equation}
\begin{eqnarray}
V&=&\frac{ \phi_{2\eta}^2}{\lambda (\phi_1 (\xi,t)+ \phi_2 (\eta,t))^2}-\frac{\phi_{2\eta\eta}}{\lambda (\phi_1 (\xi,t)+\phi_2 (\eta,t))}+  \frac{1}{2 \lambda F_2}\bigg[\frac{i}{(\alpha+\beta)}\bigg(F_{2t}+
\nonumber \\
& &\int F_2 \bigg[\frac{-i \phi_{2tt}+(\alpha+\beta)\phi_{2\eta\eta t}}{2 \phi_{2\eta}(\alpha+\beta)}  -  \frac{-i \phi_{2t} +(\alpha+\beta)\phi_{2\eta\eta}+i \kappa}{2 \phi_{2\eta}^2(\alpha+\beta)} \phi_{2\eta t}\bigg]d\eta \bigg)+ \nonumber \\
& &F_2 \bigg( \frac{-i \phi_{2 \eta t}+(\alpha+\beta)\phi_{2\eta\eta\eta}}{2 \phi_{2 \eta}(\alpha+\beta)} -\frac{-i \phi_{2t}+(\alpha+\beta)\phi_{2\eta\eta}+i \kappa}{2\phi_{2\eta}^2(\alpha+\beta)}\phi_{2\eta\eta} 
\nonumber \\ 
& & +\bigg[\frac{-i\phi_{2t}+(\alpha+\beta)\phi_{2\eta\eta}+i\kappa}{2 \phi_{2\eta}(\alpha+\beta)}\bigg]^2\bigg)\bigg].  \label{FS2}
\end{eqnarray} 
\begin{eqnarray}
U&=&\frac{\phi_{1\xi}^2}{\lambda ( \phi_1 (\xi,t)+ \phi_2 (\eta,t))^2}-\frac{ \phi_{1\xi\xi}}{\lambda ( \phi_1 (\xi,t)+ \phi_2 (\eta,t))}-\frac{F_2}{2 \lambda }\bigg[\frac{i}{(\alpha-\beta)}\bigg(\frac{-F_{2t}}{F_2^2}
+ \nonumber\\ 
& &\int \frac{1}{F_2} \bigg[\frac{-i \phi_{1tt}-(\alpha-\beta)\phi_{1\xi\xi t}}{-2 \phi_{1\xi}(\alpha-\beta)}  -  \frac{-i \phi_{1t}-(\alpha-\beta)\phi_{1\xi\xi}-i \kappa}{-2 \phi_{1\xi}^2(\alpha-\beta)} \phi_{1\xi t}\bigg] d\xi \bigg)-\nonumber \\
& &\frac{1}{F_2}\bigg( \frac{-i \phi_{1 \xi t}-(\alpha-\beta)\phi_{1\xi\xi\xi}}{-2 \phi_{1 \xi}(\alpha-\beta)}
 -\frac{-i \phi_{1t}-(\alpha-\beta)\phi_{1\xi\xi}-i \kappa}{-2\phi_{1\xi}^2(\alpha-\beta)}\phi_{1\xi\xi}
\nonumber \\
& &  +  \bigg[\frac{-i\phi_{1t}-(\alpha-\beta)\phi_{1\xi\xi}-i \kappa}{-2 \phi_{1\xi}(\alpha-\beta)}\bigg]^2\bigg)\bigg]. \label{FS3}
\end{eqnarray}  
Here, $\phi_1 (\xi,t), \phi_2 (\eta,t)), F_2(t)$ are arbitrary functions in the indicated variables and $\kappa$ is an integration constant. 

\section{Construction of Two-Dromion Solutions}
Using the arbitrariness present in the solution (\ref{GNLS22}), we choose $\phi_1$ and $\phi_2$ of the following form \cite{Tang02,Senthil05},
\begin{subequations}
\begin{eqnarray}
\phi_1  =  c_1 + l_1 \mbox{tanh}(a_1 \xi + n_1 t + d_1) +  l_2 \mbox{tanh}(a_2 \xi + n_2 t + d_2);  \\
\phi_2  =  m_1 \mbox{tanh} (b_1 \eta + p_1 t + r_1).
\end{eqnarray}
\end{subequations}
Substituting the above form of $\phi_1$ and $\phi_2$ in Eq.~(\ref{GNLS22}), we obtain the explicit form of two dromion solution analytically as,
\begin{equation}
|q|^2=\frac{q_1}{q_2}, \label{2dr}
\end{equation}
where $q_1=[l_1a_1\mbox{sech}^2(a_1\xi+n_1t+d_1)+l_2a_2\mbox{sech}^2(a_2\xi+n_2t+d_2)]m_1b_1\mbox{sech}^2(b_1\eta+p_1 t+r_1);q_2=\lambda [c_1+l_1 \mbox{tanh}(a_1 \xi + n_1 t + d_1) + 
 l_2 \mbox{tanh}(a_2 \xi + n_2 t + d_2)+m_1\mbox{tanh} (b_1 \eta + p_1 t + r_1)]^2$. The time evolution of the two dromions (\ref{2dr}) for the parametric choice $c_1 = 5; l_1 = 1; a_1 = 1; n_1 = -1; d_1 = 0;  l_2 = 1; a_2 = 1.5; n_2 \
= 2; d_2 = 0; m_2 = 1; m_1 = 1; b_1 = 1; p_1 = 1; r_1 = 1; \lambda = 0.02$ is shown in Fig.1.

\begin{figure}
\begin{center}
\includegraphics[width=0.45\linewidth]{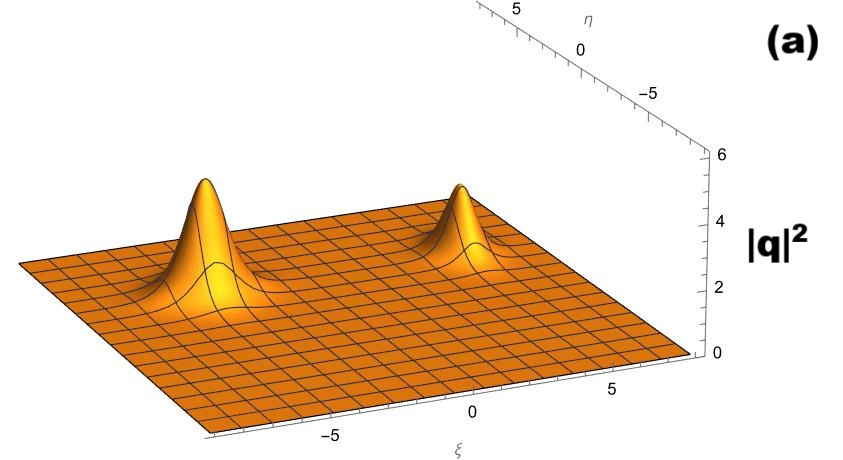}
\includegraphics[width=0.45\linewidth]{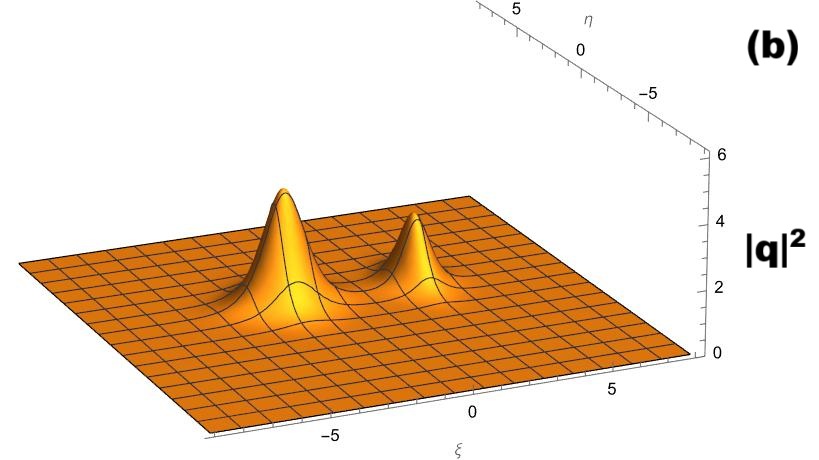}
\includegraphics[width=0.45\linewidth]{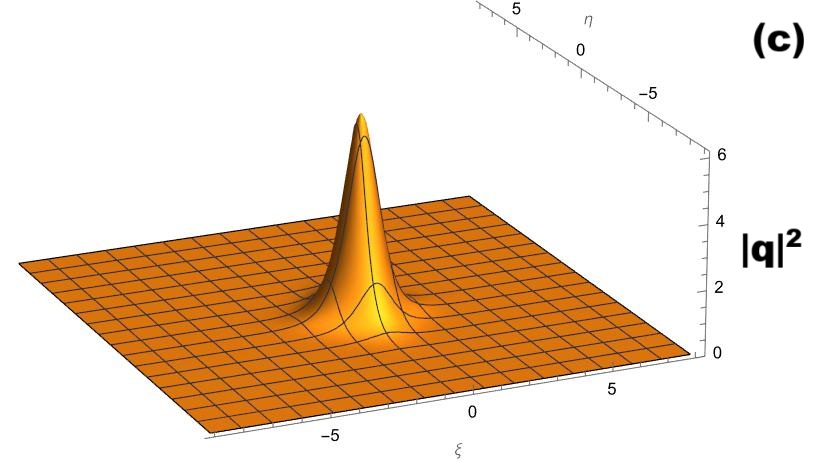}
\includegraphics[width=0.45\linewidth]{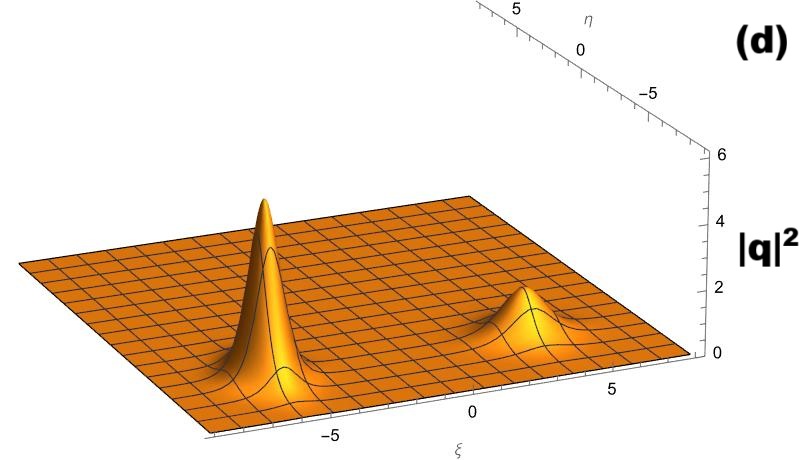}
\end{center}
\caption{Snapshots of two dromion solution at (a) t=-4 (b) t=-2 (c) t=0 (d) t=4 indicating a firewall at the boundary and reflection at the origin undergoing inelastic collision} 
\label{2drfig}
\end{figure}

\begin{figure}
\begin{center}
\includegraphics[width=0.45\linewidth]{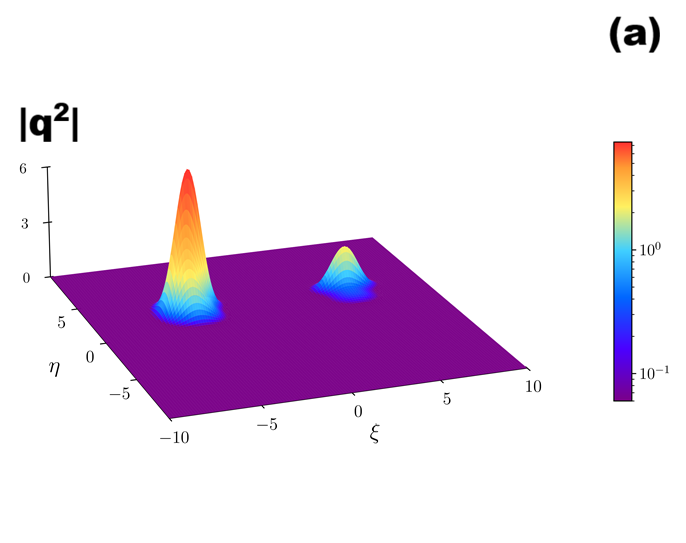}
\includegraphics[width=0.45\linewidth]{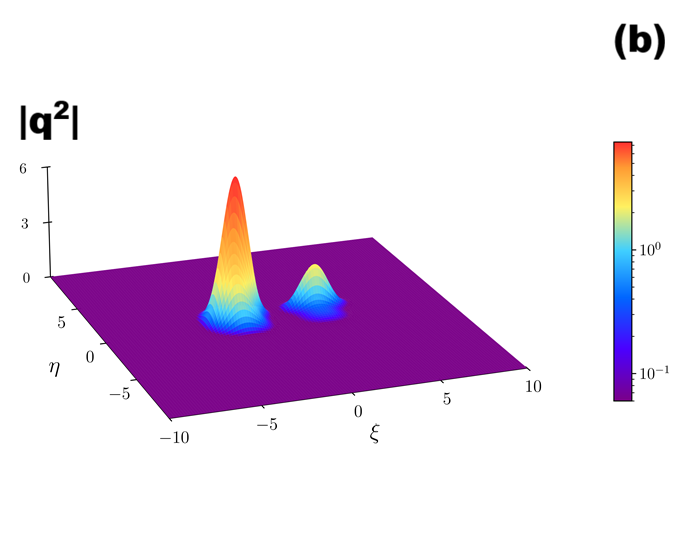}
\includegraphics[width=0.45\linewidth]{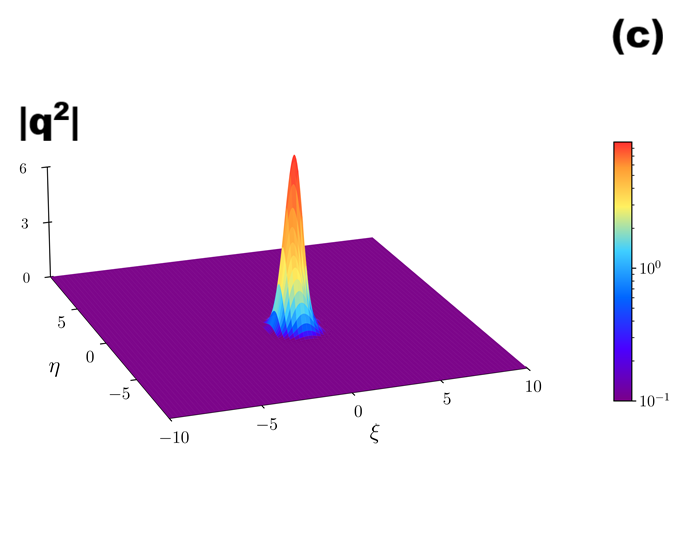}
\includegraphics[width=0.45\linewidth]{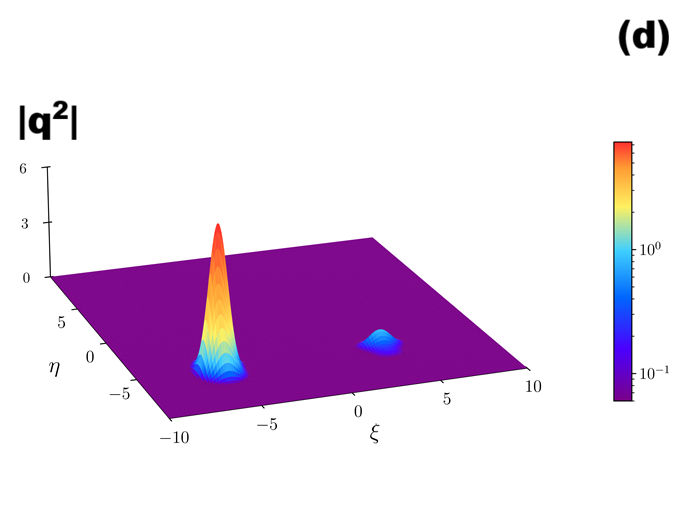}
\end{center}
\caption{ Numerically simulated density profile of two dromions at (a) t=-4 (b) t=-2 (c) t=0 (d) t=4 indicating a firewall at the boundary and reflection at the origin besides undergoing inelastic collision} 
\label{2drfignum}
\end{figure}

\section{New Insights into  Dromion Solutions}
\renewcommand{\theenumi}{\alph{enumi}}
\begin{enumerate}
\item {\bf Existence of Fire Wall and Reflection at the boundary:} In Fig.~\ref{2drfig}, we observe two dromions with differing amplitudes traveling in opposite directions. They eventually coalesce, and in the process, the shorter dromion imparts its energy to the taller dromion  implying that they undergo inelastic collision. Now, the taller dromion retraces its path in the negative $\xi$ direction. In other words, it looks as if the dromions  are prohibited in continuing their original trajectory, which implies that at the boundary, they are either reflected back or there exists a firewall which prevents the propagation of taller dromion in the positive $\xi$ direction and shorter dromion in the negative $\xi$ direction. To confirm the inelastic collision of dromions analytically, we now perform the asymptotic analysis. 
\vspace{0.4cm}

{\bf Asymptotic analysis of two dromions}
Let $u_1= a_1 \xi + n_1 t + d_1; u_2= a_2 \xi + n_2 t + d_2; u_3 = b_1 \eta + p_1 t + r_1$ and $l_1=l_2=m_1=1$  
The two dromion solution (\ref{2dr}) can then be rewritten as 
\begin{equation}
|q|^2=\frac{[a_1\mbox{sech}^2u_1+a_2\mbox{sech}^2u_2]b_1\mbox{sech}^2u_3}{\lambda [c_1+ \mbox{tanh}u_1 + 
 \mbox{tanh}u_2+\mbox{tanh}u_3]^2} \label{2dr-1}
\end{equation}
For this analysis, we set $u_3=0$ by choosing the spatial variable $\eta=0$ and the parameters $p_1=r_1=0$ in the two dromion
solution (\ref{2dr-1}) so that both the dromions travel along the $\xi$ axis.  The limit $\eta=0$ corresponds to the cross section of dromions which are essentially solitons. We analyze the limits
$t\rightarrow - \infty$  and $t\rightarrow + \infty$ separately so as to understand the interaction of dromions
centered around $u_1 \approx 0$ or $u_2 \approx 0$. Without loss of generality, let us assume
$a_1 > a_2$ and $n_1 < n_2$. Then, for the limit t $\rightarrow$ $\pm$ $\infty$, $u_1$ and $u_2$ take the following limiting values.
\newline
(1) As t $\rightarrow$ - $\infty$ :\\
$u_1 \approx 0$, $u_2 \rightarrow - \infty$,\\
$u_2 \approx 0, u_1 \rightarrow + \infty$. \\
\newline
(2) As t $\rightarrow$ + $\infty$ : \\
$u_1 \approx 0$, $u_2 \rightarrow - \infty$,\\
$u_2 \approx 0, u_1 \rightarrow + \infty$.\\

\paragraph
{1.} Before interaction (as $t \rightarrow$ -$\infty$):
For $u_1 \approx 0$, $u_2 \rightarrow -\infty$,  the two dromion
solution (\ref{2dr-1}) becomes 
\begin{equation}
	|q|^2 = \frac{a_1 b}{c_1(c_1-2)}\mbox{sech}^2(u_1+\delta_1), \delta_1=\frac{1}{2}\mbox{log}\frac{c_1}{c_1-2}
\end{equation}
For $u_2 \approx 0$, $u_1 \rightarrow  +\infty$ , the solution (\ref{2dr-1}) becomes 
\begin{equation}
	|q|^2 = \frac{a_2 b}{c_1(c_1+2)}\mbox{sech}^2(u_2+\delta_2), \delta_2=\frac{1}{2}\mbox{log}\frac{c_1+2}{c_1}
\end{equation}

\paragraph
{2.} After interaction (as $t \rightarrow$ +$\infty$):
For $u_1 \approx 0$, $u_2 \rightarrow +\infty$, the solution (\ref{2dr-1}) takes the form
\begin{equation}
	|q|^2 = \frac{a_1 b}{c_1(c_1+2)}\mbox{sech}^2(u_1+\delta_2),
\end{equation}
For $u_2 \approx 0$, $u_1 \rightarrow -\infty$, the solution (\ref{2dr-1}) becomes 
\begin{equation}
	|q|^2 = \frac{a_2 b}{c_1(c_1-2)}\mbox{sech}^2(u_2+\delta_1),
\end{equation}
From the above analysis, we observe that the amplitude of the dromions vary before and after interaction. Furthermore, the dromions undergo a phase shift, an inherent behaviour of solitons due to interaction. This is evident from the exchange of the phase factors $\delta_1$ and $\delta_2$ which are present in the argument of the respective solutions. The numerically simulated density profiles employing split-step Crank-Nicolson method shown in fig.(\ref{2drfignum}) exactly corroborate the existence of firewall and reflection at the boundary besides displaying the inelastic nature of the interaction of dromions.

\item {\bf Energy and spatial orientation of dromions} The constant term present in the sub-manifold describing the lower dimensional arbitrary function of space and time  plays a decisive role.  If it is lower than a critical value, we have a singular solution. In the two dromion solution given by Eq.~(\ref{2dr}), for example, if we choose $|c_1| \leq 3$, we obtain a  singular solution. The snapshot of the singular solution  is shown in Fig.\ref{singular}. It should also be pointed out that the constant term present in the sub-manifold imparts energy to the exponentially localized solutions thereby determining its amplitude.
\begin{figure}
\begin{center}
\includegraphics[width=0.6\linewidth]{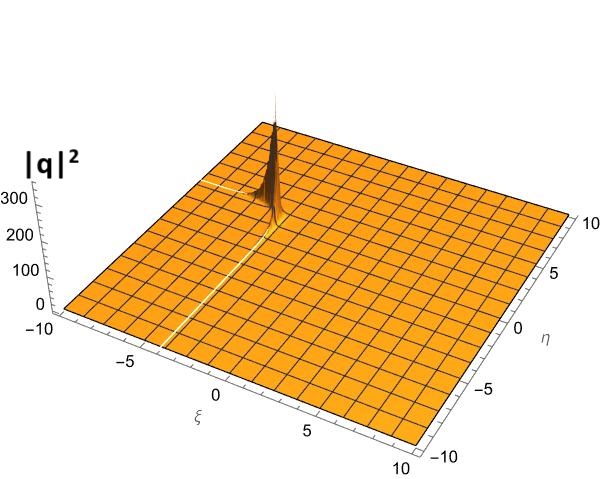}
\end{center}
\caption{Snapshot of the singular solution for the  choice $c_1 = -2; l_1 = 1; a_1 = 1; n_1 = -1; d_1 = 0;  l_2 = 1; a_2 = 1.5; n_2 
= 2; d_2 = 0; m_2 = 1; l_4 = 1; b = 1; p_1 = 1; r_1 = 1; \lambda = 1$ } 
\label{singular}
\end{figure}
 In addition to the constant parameter $c_1$,  the parameters $l_1$ and $m_1$ also control the amplitude of dromions.  The parameters $n_1$, $d_1$, $p_1$ and $r_1$ determine the spatial orientation of dromions  while the parameters $a_1$ and $b_1$ can either squeeze or disperse/spatially spread the dromions.

\item {\bf Nonuniform  Distribution of Energy:} It should be pointed out that  the energy is not evenly distributed among the dromions. Let us consider higher order dromions. For example, let us consider a train of seven ghost solitons in one spatial direction and  another ghost soliton in the other spatial direction by choosing $\phi_1$ and $\phi_2$ as 
\begin{subequations}
\begin{eqnarray}
\phi_1 &=& c_1 + l_1 \mbox{tanh}(a_1 \xi + n_1 t + d_1)+
 l_2 \mbox{tanh}(a_2 \xi + n_2 t + d_2)+ \nonumber \\
  & &  l_3 \mbox{tanh}(a_3 \xi + n_3 t + d_3)+ 
l_4 \mbox{tanh}(a_4 \xi + n_4 t + d_4)+ \nonumber \\
& &  l_5 \mbox{tanh}(a_5 \xi + n_5 t + d_5)+ l_6 \mbox{tanh}(a_6 \xi + n_6 t + d_6)+ \nonumber \\
&& 
 l_7 \mbox{tanh}(a_7 \xi + n_7 t + d_7),  \\
 \phi_2 &=& m_1 \mbox{tanh} (b_1 \eta + p_1 t + r_1).
\end{eqnarray} \label{7dr}
\end{subequations}
Substituting Eq.~(\ref{7dr}) containing  $\phi_1$ and $\phi_2$ in Eq.~(\ref{GNLS22}),  we get 
\begin{equation}
|q|^2=\frac{q_3}{q_4}
\end{equation}
where 
\begin{eqnarray}
    q_3 & = & b_1 m_1 \mbox{sech} (b_1 \eta + p_1 t + r_1)^2 
  [a_1 l_1 \mbox{sech}(a_1 \xi + n_1 t + d_1)^2   \nonumber \\ 
& & 
+ a_2 l_2 \mbox{sech}(a_2 \xi + n_2 t + d_2)^2 + 
   a_3 l_3 \mbox{sech}(a_3 \xi + n_3 t + d_3)^2 +  \nonumber \\ 
& & 
     a_4 l_4 \mbox{sech}(a_4 \xi + n_4 t + d_4)^2 + 
   a_5 l_5 \mbox{sech}(a_5 \xi + n_5 t + d_5)^2 +
   \nonumber \\
& &   a_6 l_6 \mbox{sech}(a_6 \xi + n_6 t + d_6)^2 + 
   a_7 l_7 \mbox{sech}(a_7 \xi + n_7 t + d_7)^2] \nonumber,
\end{eqnarray}

   and 
\begin{eqnarray}
    q_4 &=& \lambda [c_1 + l_1 \mbox{tanh}(a_1 \xi + n_1 t + d_1) + 
 l_2 \mbox{tanh}(a_2 \xi + n_2 t + d_2)+  \nonumber \\ 
& & 
 l_3 \mbox{tanh}(a_3 \xi + n_3 t + d_3)+ 
 l_4 \mbox{tanh}(a_4 \xi + n_4 t + d_4)+ 
  \nonumber \\
& & 
 l_5 \mbox{tanh}(a_5 \xi + n_5 t + d_5)+ 
 l_6 \mbox{tanh}(a_6 \xi + n_6 t + d_6)+ \nonumber \\
& & 
 l_7 \mbox{tanh}(a_7 \xi + n_7 t + d_7)+m_1 \mbox{tanh} (b_1 \eta + p_1 t + r_1)]^2  \nonumber.
\end{eqnarray}  
 Giving equal input through the parameters $l_i $s and $m_i $s, we obtain a  seven dromion solution whose  snapshot  is shown in   Fig.~\ref{Uneven}.  From the figure, it is obvious that the seven dromions have different amplitudes.   In other words, there is an uneven distribution of energy among dromions. This property is similar to the Fermi-Pasta-Ulam problem, where the energy supplied is shared among the first few modes violating equipartition of energy \cite{Lakshmanan03}.  Instead of  giving  equal input to all the ghost solitons through parameters $l_i s$ and $m_i s$ to generate  dromions, we notice that even if one of the bound states is given a different input through parameter $l_i$, the energy distribution is  altered as shown in  Fig.~\ref{7dr equal input except l5}. Comparing Figs.~\ref{Uneven} and \ref{7dr equal input except l5}, one observes that the energy keeps rising gradually among the seven bound states, while the sequence of distribution of energy is disturbed by imparting an unequal input through parameter $l_5$ thereby changing the landscape of distribution of energy.

\begin{figure}
\begin{center}
\includegraphics[width=0.7\linewidth]{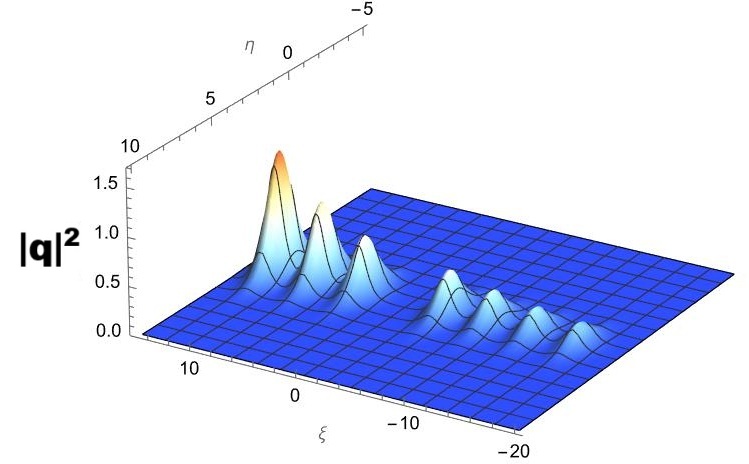}
\end{center}
\caption{Uneven distribution of energy of dromions by giving  equal inputs to all the bound states, $l_1 = 1; a_1 = 1; n_1 = 1; d_1 = 0; l_2 = 1; a_2 = 1; n_2 = 2; d_2 = 0; l_3 = \
1; a_3 = 1; n_3 = 3; d_3 = 0; l_4 = 1; a_4 = 1; n_4 = -1; d_4 = 0; l_5 = 1; a_5 = 1; n_5 = -2; d_5 = 0; l_6 = 1; a_6 = 1; n_6 = -3; d_6 = 0; l_7 = 1; a_7 = 1; n_7 = -4; d_7 = 0; k_1 = 1; m_1=1; b_1 = 1; p_1 = 1; r_1 = 1; \lambda = 0.01; t = -4; c_1 = -15$ } 
\label{Uneven}
\end{figure}

\begin{figure}
\begin{center}
\includegraphics[width=0.7\linewidth]{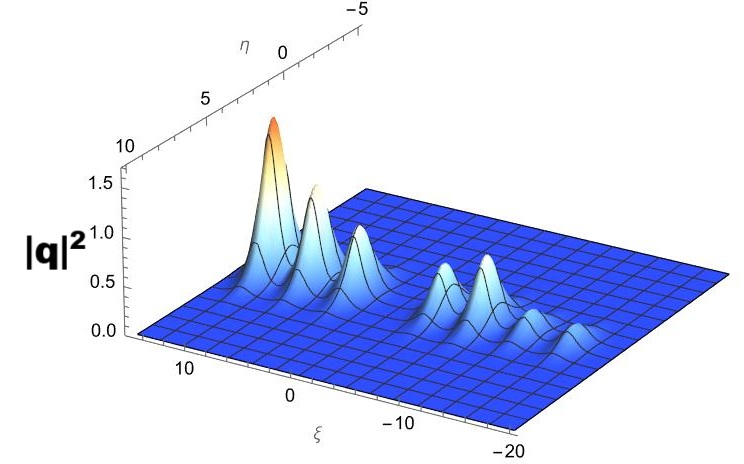}
\end{center}
\caption{Energy distribution by giving  equal inputs except for one bound state, $l_1 = 1; a_1 = 1; n_1 = 1; d_1 = 0; l_2 = 1; a_2 = 1; n_2 = 2; d_2 = 0; l_3 = \
1; a_3 = 1; n_3 = 3; d_3 = 0; l_4 = 1; a_4 = 1; n_4 = -1; d_4 = 0; l_5 = 2; a_5 = 1; n_5 = -2; d_5 = 0; l_6 = 1; a_6 = 1; n_6 = -3; d_6 = 0; l_7 = 1; a_7 = 1; n_7 = -4; d_7 = 0; k_1 = 1; m_1=1; b_1 = 1; p_1 = 1; r_1 = 1; \lambda = 0.01; t = -4; c_1 = -15$ } 
\label{7dr equal input except l5}
\end{figure}

\item {\bf Sequence of Distribution of Energy:} Again, considering  the seven dromion solution as shown in Fig.~\ref{Uneven}, we wish to point out that it is associated with another interesting property. In Fig.~\ref{Uneven}, the amplitude of the dromions is decreasing in a descending order. In other words, in a multidromion solution,  the distribution of energy can be  either in ascending or descending order as shown in Fig.~\ref{Uneven} and \ref{order of distribution}.
 This is controlled by the sign of the constant parameter. For example, in the case of seven dromions, if we choose $c_1=-15$, we get the array of dromions as shown in Fig.~\ref{Uneven} where the energy of the bound states keeps on decreasing in a descending order. If we choose $c_1=15$, we get a different pattern as shown in Fig.~\ref{order of distribution}, where the energy keeps on increasing.
\begin{figure}
\begin{center}
\includegraphics[width=0.7\linewidth]{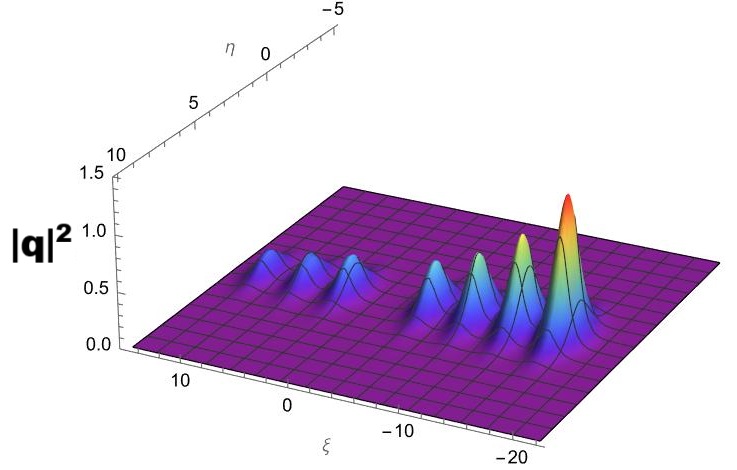}
\end{center}
\caption{Seven dromion solution with increasing amplitude (ascending order)  for the parametric choice, $l_1 = 1; a_1 = 1; n_1 = 1; d_1 = 0; l_2 = 1; a_2 = 1; n_2 = 2; d_2 = 0; l_3 = \
1; a_3 = 1; n_3 = 3; d_3 = 0; l_4 = 1; a_4 = 1; n_4 = -1; d_4 = 0; l_5 = 1; a_5 = 1; n_5 = -2; d_5 = 0; l_6 = 1; a_6 = 1; n_6 = -3; d_6 = 0; l_7 = 1; a_7 = 1; n_7 = -4; d_7 = 0; k_1 = 1; m_1=1; b_1 = 1; p_1 = 1; r_1 = 1; \lambda = 0.01; t = -4; c_1 = 15$
} 
\label{order of distribution}
\end{figure}

\item {\bf Amplitude dependence on neighbouring dromions} The presence or absence of dromions is found to control the amplitude of the neighbouring dromions.  For example, in Fig.~\ref{amplitude dependence on neighbours}a, there are seven dromions with their amplitudes arranged in an ascending order. 
In the seven dromion solution given by the arbitrary function $\phi_1$ given in Eq.~\ref{7dr}a, we nullify one of the dromions  by choosing the parameter $l_1=0$.  The corresponding plot is shown in Fig.~\ref{amplitude dependence on neighbours}b. Now, we make one more dromion disappear by choosing  $l_2=0$ and  the corresponding plot is shown in Fig.~\ref{amplitude dependence on neighbours}c. Similarly three dromions are nullified by choosing $l_1=l_2=l_3=0$ and the corresponding profile is shown in Fig.~\ref{amplitude dependence on neighbours}d and four dromions made zero by choosing $l_1=l_2=l_3=l_4=0$ which is depicted Fig.~\ref{amplitude dependence on neighbours}e.

\begin{figure}
\begin{center}
\includegraphics[width=0.45\linewidth]{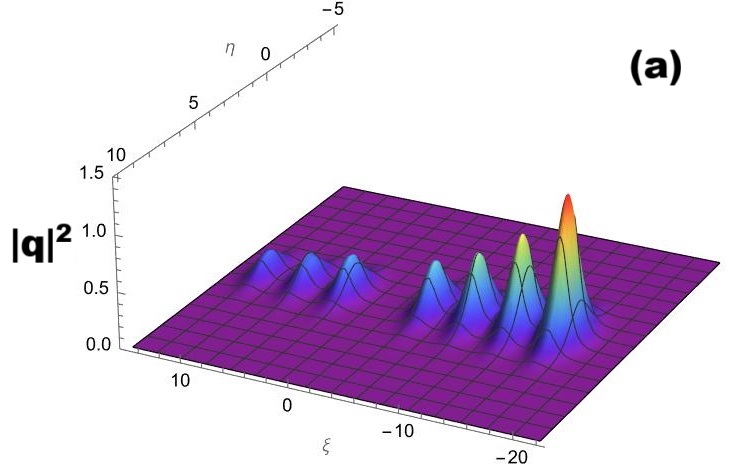}
\includegraphics[width=0.45\linewidth]{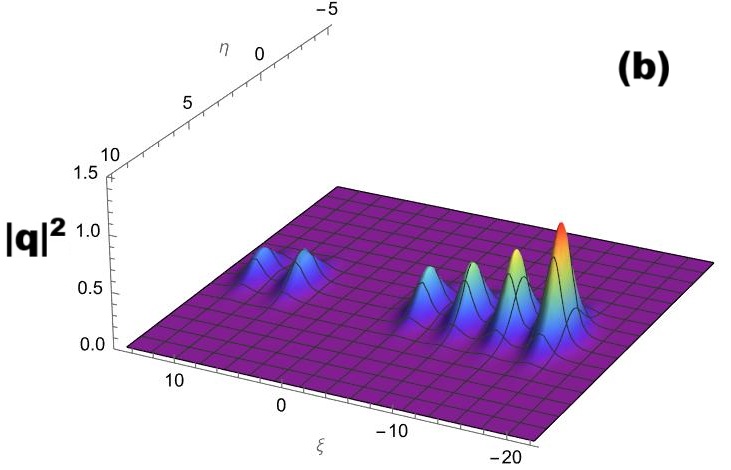}
\includegraphics[width=0.45\linewidth]{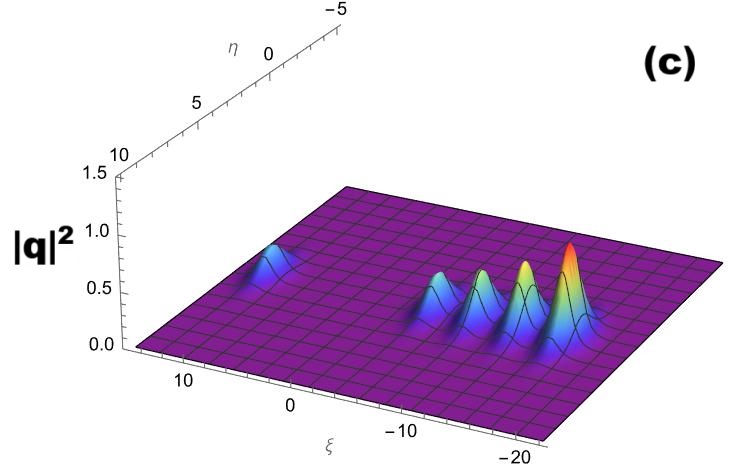}
\includegraphics[width=0.45\linewidth]{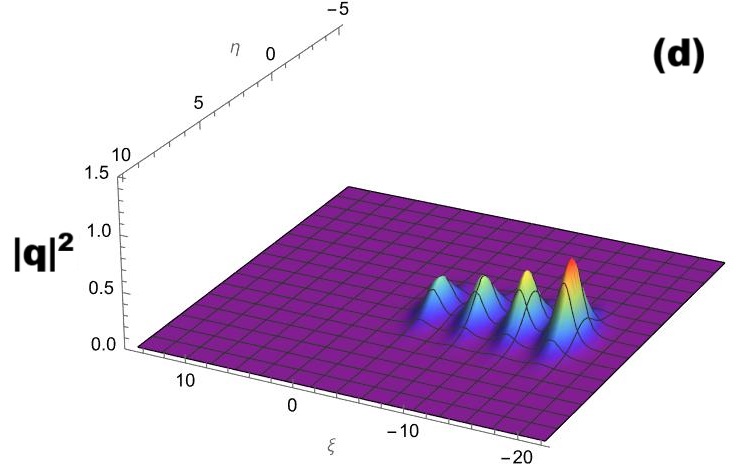}
\includegraphics[width=0.45\linewidth]{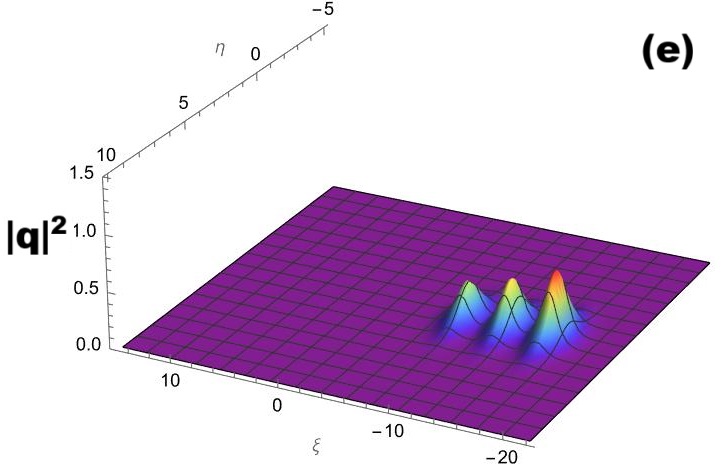}
\end{center}
\caption{Snapshots of the seven dromion solution with seven bound states  at  t=-4 (a) all the seven dromions present (b) one dromion is nullified (c) two dromions are nullified (d) three dromions are nullified  (e) four dromions are nullified  
} 
\label{amplitude dependence on neighbours}
\end{figure}

\begin{figure}
\begin{center}
\includegraphics[width=0.45\linewidth]{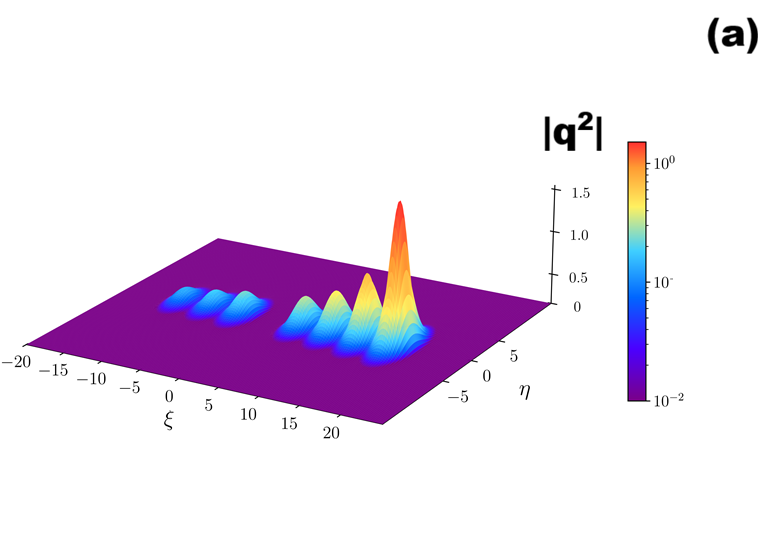}
\includegraphics[width=0.45\linewidth]{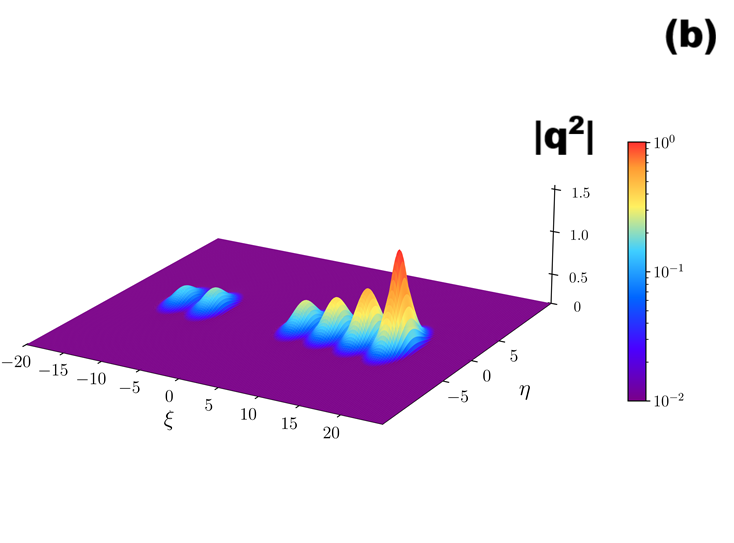}
\includegraphics[width=0.45\linewidth]{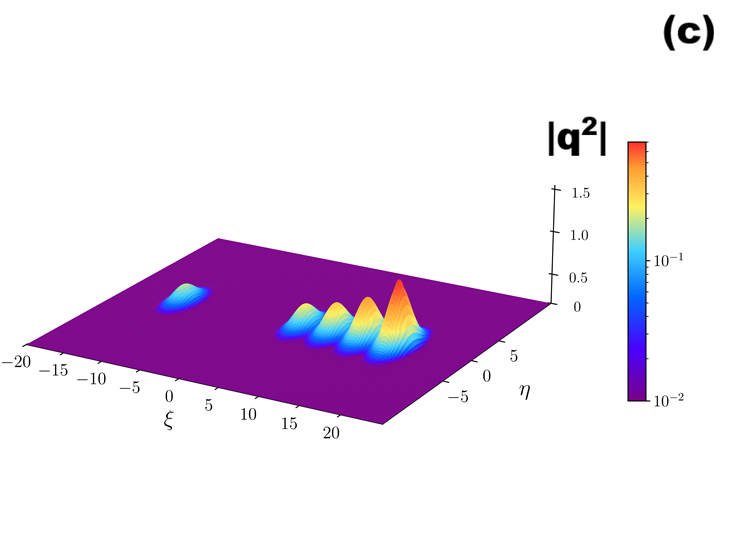}
\includegraphics[width=0.45\linewidth]{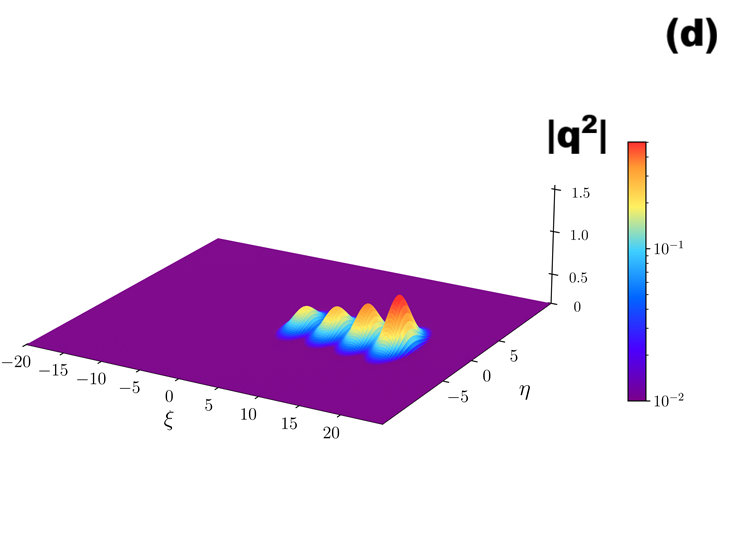}
\includegraphics[width=0.45\linewidth]{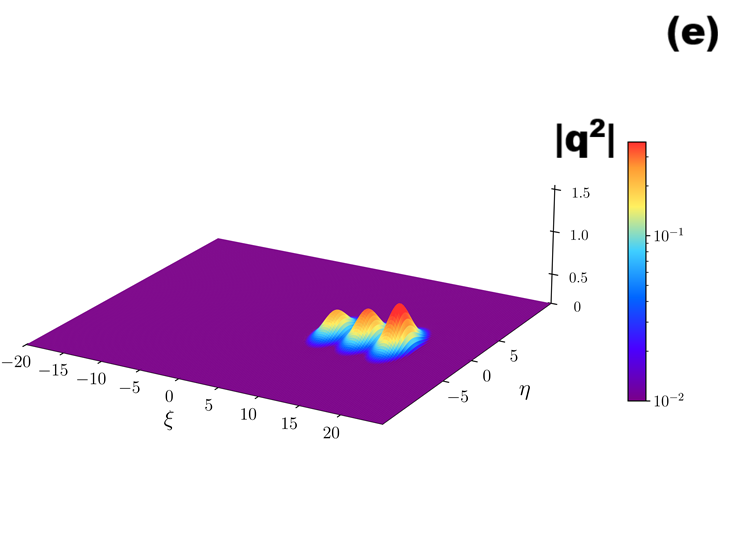}
\end{center}
\caption{Numerically simulated density profile of the seven dromion solution with seven bound states  at  t=-4 (a) all the seven dromions present (b) one dromion is nullified (c) two dromions are nullified (d) three dromions are nullified  (e) four dromions are nullified} 
\label{numerical amplitude dependence on neighbours}
\end{figure}

By comparing the plots shown in Fig.~(\ref{amplitude dependence on neighbours}), it is obvious that there is a decrease in the amplitude of the  dromion cluster  from Fig.~\ref{amplitude dependence on neighbours} a - \ref{amplitude dependence on neighbours} e. Thus, the presence or absence of a  dromion is impacting the amplitude of the adjacent dromions. This  dependence of the dromions's amplitude on that of the neighbours is also comfirmed numerically in Fig.(\ref{numerical amplitude dependence on neighbours}).  In other words, the dromions are nonlinearly coupled to each other and hence  each  dromion can influence the other dromions in its viscinity either allowing them to vanish or gain amplitude. 

This property can also be verified in (1+1) dimensional nonlinear systems. For example, we consider the two soliton solution of the K-dV equation \cite{Lakshmanan03} as shown in  Fig.~\ref{kdv 2 soliton}. If we remove one bound state by choosing the spectral parameter $k_1=0$, the amplitude of the remaining soliton gets reduced as shown in Fig.~\ref{kdv 2 soliton one spec parm zero}. In addition, there is a marginal phase shift occurring in the other bound state(soliton). This again confirms the nonlinear coupling of the solitons.

\begin{figure}
\begin{center}
\includegraphics[width=0.7\linewidth]{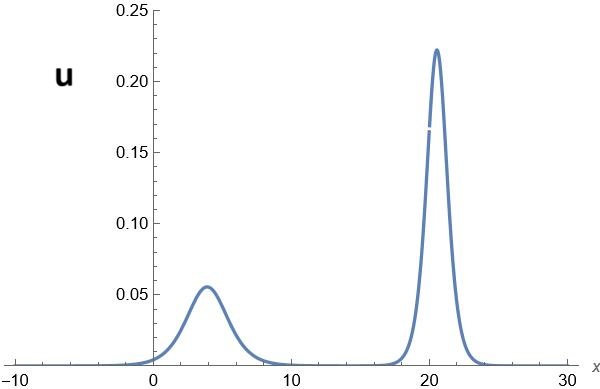}
\end{center}
\caption{Two soliton solution of K-dV equation with two bound states (spectral parameters)
} 
\label{kdv 2 soliton}
\end{figure}

\begin{figure}
\begin{center}
\includegraphics[width=0.7\linewidth]{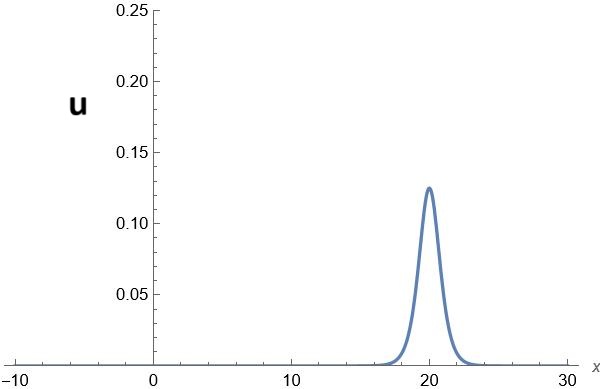}
\end{center}
\caption{Two soliton solution of K-dV equation by removing one bound state (by choosing the spectral parameter $k_1=0$)
} 
\label{kdv 2 soliton one spec parm zero}
\end{figure}
From the above discussion, it is clear that each dromion exerts a strong influence on the surrounding dromions.
\end{enumerate}

All the above properties are verified in other (2+1) dimensional nonlinear systems and the results are tabulated in Table (\ref{comparitive}). The $\checkmark$ symbol indicates that the corresponding property is exhibited by the respective system. From the tabulated results, we find that all the systems tabulated exhibit all the novel properties reported in this paper thereby, showing the universality.  It must be mentioned that the first property, i.e. existence of fire wall exhibited by the KdV and AKNS system is valid for a specific parametric choice \cite{SENTHILKUMAR2009942,RADHA20182356}.

\begin{table}[]
    \centering
           \begin{tabular}{|c|c|c|c|c|c|c|}
    \hline
         S. No. &  Dynamical Systems 
         & 
         \begin{turn}{90} {\begin{tabular}{c} Property 1: \\ Existence of Fire Wall and \\ Reflection at the boundary \end{tabular}} \end{turn} 
         & 
          \begin{turn}{90} {\begin{tabular}{c} Property 2: \\ Energy and spatial \\ orientation of dromions \end{tabular}} \end{turn}
         &  
         \begin{turn}{90} {\begin{tabular}{c} Property 3: \\ Non uniform \\ Distribution of Energy
         \end{tabular}} \end{turn}
         & 
         \begin{turn}{90} {\begin{tabular}{c} Property 4: \\ Sequence of \\Distribution of Energy
         \end{tabular}} \end{turn}
         & 
         \begin{turn}{90} {\begin{tabular}{c} Property 5: \\ Amplitude dependence \\ on neighbouring dromions
         \end{tabular}} \end{turn}\\
         \hline
         1 & {\begin{tabular}{c} Generalized NLS \\ (DSI and DS III, \\ (2+1) Scalar NLS)   \end{tabular}} & \checkmark & \checkmark & \checkmark & \checkmark & \checkmark \\ \hline
         2 & {\begin{tabular}{c} Long Wave Short Wave \\ Resonance Interaction \\ (LSRI) equation  \cite{Senthil05}\end{tabular}}  & \checkmark & \checkmark & \checkmark & \checkmark & \checkmark \\ \hline
         3 & {\begin{tabular}{c} Ablowitz-Kaup-Newell- \\Segur (AKNS) equation \end{tabular}} & \checkmark & \checkmark & \checkmark & \checkmark & \checkmark \\ \hline
         4 & {\begin{tabular}{c} Nizhnik–Novikov– \\Veselov (NNV) equation \end{tabular}} & \checkmark & \checkmark & \checkmark & \checkmark & \checkmark \\ \hline
         5 & {\begin{tabular}{c} (2+1) dimensional  \\ Korteweg-de Vries \\(KdV) equation \end{tabular}} & \checkmark & \checkmark & \checkmark & \checkmark & \checkmark \\ \hline
         6 & {\begin{tabular}{c} Boiti-Leon-Pempinelli \\ (BLP) equation \end{tabular}}  & \checkmark & \checkmark & \checkmark & \checkmark & \checkmark \\ \hline
         7 & {\begin{tabular}{c} Broer-Kaup- \\Kupershmidt \\ (BKK) equation \end{tabular}} & \checkmark & \checkmark & \checkmark & \checkmark & \checkmark \\ \hline
         8 & {\begin{tabular}{c} (2+1) dimensional \\ long  dispersive wave \\ (LDW) equation \end{tabular}} & \checkmark & \checkmark & \checkmark & \checkmark & \checkmark \\ \hline
         9 & {\begin{tabular}{c}(2+1) dimensional \\ Sine-Gordon \end{tabular}} & \checkmark & \checkmark & \checkmark & \checkmark & \checkmark \\
         \hline
    \end{tabular}
    \caption{An Overview of the Validity  of the Properties in  2+1 Dimensional Nonlinear Systems}
    \label{comparitive}
     \end{table}

\section{Conclusion}
In this paper, we have employed the Truncated Painlev\'{e} Approach for generating the localized solutions of the (2+1) dimensional generalized NLS Equation. We have generated its solution in closed form in terms of lower dimensional arbitrary functions of space and time.  We have then constructed its multi dromion solutions and studied their collisional dynamics to bring out their exotic behaviour which have never been reported so far. 
To bring out its universality, we have verified this in other (2+1) dimensional nonlinear pdes and tabulated at the end.

It would be interesting to explore whether dromions exhibit other peculiar properties other than the ones reported here. It should also be pointed out that even though the Painlev\'e Truncation approach has been exploited to construct the  closed form solutions of (2+1) dimensional nonlinear pdes with lower dimensional arbitrary functions of space and time, it does have its limitations.  It can be applied only to  (2+1) dimensional completely integrable nonlinear pdes alone whereas one comes across several nonintegrable (2+1) dimensional nonlinear pdes admitting light bullets \cite{silberberg1990solitons}, skyrmions \cite{roessler2006spontaneous}  which can be realized experimentally for which closed form solutions have not yet been extracted.

\section{Acknowledgements}
CSK wishes to thank Vinayaka Mission's Research Foundation (Deemed to be University) for financial support in the form of seed money  (VMRF/Research/ Seed Money/2022-23/VMKVEC-08). RR wishes to acknowledge financial assistance received from DST-CURIE(DST-CURIE/PG/54/2022) and ANRF(DST-CRG/ 008153/2023) in the form of sponsored research projects. We thank the referees for their valuable suggestions in improving the contents of the manuscript.

\newpage
\bibliographystyle{elsarticle-num-names}
\bibliography{apssamp}

\end{document}